\documentclass[fleqn,usenatbib,referee]{mnras}

\usepackage{newtxtext,newtxmath}

\usepackage[T1]{fontenc}
\usepackage{ae,aecompl}

\usepackage{graphicx}	
\usepackage{amsmath}	
\usepackage[utf8]{inputenc}
\usepackage[british]{babel}
\usepackage{siunitx}
\usepackage{multicol}
\usepackage{multirow}
\usepackage{subfigure}
\usepackage{booktabs}
\usepackage{wasysym}
\usepackage{epstopdf}

\title[Can a jumping-Jupiter trigger the Moon's formation impact?]{Can a jumping-Jupiter trigger the Moon's formation impact?}

\author[S. R. DeSouza et al.]{
Sandro R. DeSouza,$^{1}$
Fernando Roig,$^{1}$\thanks{E-mail: froig@on.br}
David Nesvorn\'{y}$^{2}$
\\
$^{1}$Observat\'{o}rio Nacional, ON, Rio de Janeiro, 20921-400, RJ, Brazil\\
$^{2}$Southwest Research Institute, SWRI, Boulder, CO 80302, USA
}

\date{Accepted XXX. Received YYY; in original form ZZZ}

\pubyear{2021}

\begin{document}
\label{firstpage}
\pagerange{\pageref{firstpage}--\pageref{lastpage}}
\maketitle

\begin{abstract}
We investigate the possibility that the Moon's formation impact was triggered by an early dynamical instability of the giant planets. We consider the well-studied “jumping Jupiter” hypothesis for the solar system’s instability, where Jupiter and Saturn’s semi-major axes evolve in step-wise manner from their primordially compact architecture to their present locations. Moreover, we test multiple different configurations for the primordial system of terrestrial planets and the Moon-forming projectile, with particular focus on the almost equal masses impact.
We find that the instability/migration of the giant planets 
excites the orbits of the terrestrial planets through dynamical perturbations, thus allowing collisions between them.
About $10$\% of the simulations lead to a collision with the proto-Earth which resulted in a final configuration of the terrestrial system that reproduces, to some extent, its present architecture. 
Most of these collisions occur in the hit-and-run domain, but about 15\% occur in the partial accretion regime, with the right conditions for a Moon-forming impact.
In most of the simulations, there is a delay of more than $\sim \num{20}\,\si{My}$ between the time of the instability and the Moon-forming impact. This supports the occurrence of an early instability ($< \num{10}\,\si{My}$ after dissipation of the gas in the proto-planetary disk), compatible with the time of the Moon-forming impact (30-$60\,\si{My}$) inferred from cosmochemical constraints.
In general, the final states of the inner solar system in our simulations show an excess of Angular Momentum Deficit, mostly attributed to the over-excitation of Mercury's eccentricity and inclination.
\end{abstract}



\section{Introduction}

The standard model of terrestrial planets formation can be divided into three stages: (i) the formation of planetesimals from dust grains near the mid-plane of the proto-planetary nebula by streaming instability \citep[e.g.][]{Morby2009}, (ii) the runaway growth of embryos, in which small proto-planets form in circular and co-planar orbits \citep[e.g.][]{Kokubo1996,Clement2020}, and (iii) the aggregation of the embryos that collide due to their perturbations on the neighbours, thus forming the terrestrial planets \citep[e.g.][]{Chambers1998}. In this context of planetary formation, it is accepted that the Moon formed from the impact of a Mars-sized or larger proto-planet on the proto-Earth, that would have occurred during the third stage. In this ``Giant Impact'' model, the Moon accretes from the material ejected by the collision. Isotopic dating of Earth core formation indicates that this event happened 30 to 60 My after the gas dispersal in the proto-planetary disk, some $\num{4.56}$ billion years ago \citep[e.g.][]{Barboni2017}, although the precise timing is still a matter of debate.

The analysis of isotopic abundances in the lunar rocks provides strong constraints on the Giant Impact model. In the standard model, with a Mars-size impactor, it is expected that the chemical signatures of the Moon reproduce to some degree those of the impactor. Measurements of isotopic ratios of different elements have shown that Mars is different from the Earth \citep[e.g.][]{Georg2007}, implying a significant gradient of isotopic abundances between 1.0 and 1.5 au. However, the Earth and the Moon are almost identical in terms of isotopic abundances \citep[e.g.][]{Wiechert2001}. These results would be consistent with the Giant Impact hypothesis if the proto-Earth and the impactor were formed from an identical mixture of components, at approximately the same heliocentric distance. Other explanations suggest, instead, that the Moon’s isotopic composition reflects only the Earth's contribution \citep[e.g.][]{Zhang2012,Cuk2018}, or that two half-Earth-like proto-planets collided forming the Earth-Moon pair and getting the correct balance in the isotopic composition \citep{Canup2012}, or even that the isotopes equilibrated or mixed during the vapour phase of the proto-lunar disk. 

The formation of the Moon has been addressed by several previous studies, using different models and testing different configurations: hit-and-run collision \citep[e.g.][]{Reufer2012}, equal mass collision \citep[e.g.][]{Canup2012}, fast-spinning target collision \citep[e.g.][]{Cuk2012}, collision within N-body accretion simulations \citep[e.g.][]{Kaib-Cowan2015}, studies on the pre-impact orbital configurations \citep[e.g.][]{Quarles2015,Jackson2018}, among others; see \citet{Canup2014} for a review.

We are interested here in the specific case of almost equal mass collision. This model has the advantage that the resulting Earth-Moon system may share the same composition, because the impactor does not only contribute to the Moon's accretion disk, but also significantly affects the composition of the target.\footnote{In a canonical impact (projectile-to-target ratio $\sim 1/10$), significant removal of mass from the target is necessary to equalise the composition of the final system.}
On the other hand, equal mass impacts leave the final system with an excess of angular momentum that has to be removed later through other dynamical processes, like evection resonance interaction \citep[e.g.][]{Cuk2012}.

\citet{Canup2012} explored the nearly equal mass model through hydrodynamics simulations, considering two bodies with approximately half Earth mass each. She found that Moon-forming collisions may arise over a wide range of impact velocities and angles. The main limitation of this model, however, is that the actual likelihood of collisions between similar-sized planets in standard accretion models might be low. Another problem is the initial orbital location of the colliding pair. \citet{Quarles2015}, for example, determined that the progenitors need to be quite close in terms of semi-major axis.
Finally, a collision with a half-Earth impactor may have the potential to significantly excite the Earth's orbit, although \citet{Canup2012} found that 70\% of her successful simulations produce orbital velocity changes, which would not cause any significant excitation.

In this work, we investigate the conditions that may lead to the collision of two initially proto-planets in close radial proximity in the inner solar system. Specifically, we want to address the role of the giant planets instability in triggering this event. The instability of the giant planets has became a fundamental ingredient of the primordial dynamical evolution of the solar system, particularly in the context of the Nice model \citep{Tsiganis2005,Morbidelli2005,Gomes2005}. This model assumes that, after dissipation of the gas in the proto-planetary disk, the recently formed giant planets started to gravitationally interact with a disk of remnant planetesimals beyond the orbit of Neptune. This interaction drove the radial migration of the giant planets and could have favoured the occurrence of mutual mean motion resonances and close encounters between them, eventually leading to a temporary instability of the system. Such an instability had a major role in sculpting many of the characteristics that we observe nowadays in the solar system \citep[see][for a review]{Nesvorny2018}.
In particular, \citet{Clement2020} proposed that the instability might be needed to destabilise a primordial system of more massive terrestrial embryos, and trigger the Moon forming impact.

Of particular interest is the type of instability known as the jumping Jupiter model \citep{Brasser2009,Morbidelli2009}. In this case, a close encounter between Jupiter and an ice giant caused an exchange of angular momentum that led to the ejection of the ice giant from the system while Jupiter migrated inwards in a discontinuous and abrupt way, virtually ``jumping'' to an inner orbit. This model has proved to be very successful to address several dynamical properties of the solar system bodies like Main Belt asteroids \citep[][]{Morbidelli2015,Roig2015,BRASIL2016,Brasil2017}, Trojan asteroids \citep[][]{Nesvorny2013}, trans-Neptunian objects \citep[][]{Nesvorny2015}, terrestrial planets \citep[][]{Roig2016}, satellites of the giant planets \citep[][]{Nesvorny2014,Deienno2014}, to cite a few.

The jumping Jupiter instability is relevant for the formation and early evolution of the terrestrial planets.
A smooth migration of the giant planets causes a slow sweeping of secular resonances through the inner solar system, that contributes to destabilise the terrestrial planets. The jumping Jupiter, on the other hand, swiftly moves the secular resonances to their present location, preventing orbits in the terrestrial region from becoming overly excited \citep[e.g.][]{Brasser2009,Brasser2013,Agnor2012,Roig2016}. 

However, the time of the instability is still a matter of debate. A late instability $(t\gtrsim 100\,\si{My})$ has been the preferred scenario since, in principle, it would help to justify the lunar Late Heavy Bombardment. On the other hand, recent developments favour the occurrence of an early instability  \citep[$t<10$\,\si{My};][]{Kaib2016,Nesvorny2018NatAs,Morbidelli2018}, and raise questions on how this could affect the terrestrial planet formation \citep[e.g.][]{Clement2018,Clement2019}. Here, we focus on the specific problem of the Moon-forming collision in the framework of an early instability. 

The late stages of the standard model of terrestrial planets accretion, where the proto-planets were on crossing orbits, could provide the natural conditions for the Giant Impact to occur. Here, instead, we propose that the Giant Impact could have been triggered by an early jumping Jupiter instability, that favoured the collision of two proto-planets formed at around 1 au and initially separated by $\sim 0.1\,\si{au}$.
To test this hypothesis, we assume that the terrestrial planets are already formed (or almost formed), and they reside in cold orbits (very low eccentricity and inclination) when the instability occurs. This initial setup is motivated by three reasons:
\begin{enumerate}
\item We want to schematise things and isolate the effect of the instability to properly quantify it. In this sense, we may suppose to be working with a sort of toy model, which does not necessarily reflect the real conditions of the system when the instability happens.
\item There are alternative models of terrestrial planets formation showing that the planets may have formed within a gas disk \citep[e.g.][]{Ogihara2018}. In particular, \cite{Broz2021} invoke torque-driven convergent migration of planetary embryos in the proto-planetary gas disk to show that terrestrial planets may have accreted much faster than in the standard model. Measurements of isotopic anomalies \citep{Dauphas2011,Schiller2020} also support the fast formation of terrestrial proto-planets, possibly while the gas disk was still around. In this scenario, the proto-terrestrial planets would have emerged from the gas disk nearly formed and on non-crossing orbits. Therefore, in order to have the Giant Impact at 30-60\,\si{My}, it is necessary to have a mechanism to destabilise the orbits, and this might be provided by the jumping Jupiter evolution.
\item High-resolution studies of runaway growth of terrestrial planets show that it would be very efficient mechanism of accretion, and it is very possible that a quasi-stable system of larger ($\sim 0.3$-0.5\,\si{M_\oplus}) proto-planets existed for some time after nebular gas dispersal \citep{Walsh2019,Clement2020}.
\end{enumerate}
In particular, we focus on the analysis of a configuration with a projectile-to-target mass ratio of $\sim 1$, considering the presence in the system of two proto-planets, each with about half the mass of the Earth \citep{Canup2012}. We also analyse in less detail a configuration with a projectile-to-target mass ratio of $\sim 1/10$, in which we have an additional Mars-like proto-planet in the system. 

\begin{table*}
  \centering
  \caption[Initial orbital elements and masses of the planets]{Initial orbital elements and masses of the planets in the different models. The elements of the primordial giants are those at the beginning of Phase 1. The elements of the present giants correspond to JD 2451544.5. $\Delta a$ may be 0.05, 0.08, 0.1, 0.15 or 0.2 au.}%
  \label{tab:initial_conditions}%
    \begin{tabular}{lccccc}%
    \toprule
     \multicolumn{1}{l}{Configuration} & \multicolumn{1}{c}{Planets} & \multicolumn{1}{c}{$\text{Mass}\,(M_{\text{Jup}})$} & \multicolumn{1}{c}{$a\,(\si{au})$} & \multicolumn{1}{c}{$e$} & \multicolumn{1}{c}{$I\,(\si{\degree})$} \\
    \midrule
    \multirow{3}{*}{\shortstack[l]{Common}} 
        & Mercury   & \num{0.00017} & \num{0.387}   & \num{0.001} & \num{0.01} \\
		& Venus     & \num{0.00256} & \num{0.723}   & \num{0.001} & \num{0.01} \\
    	& Mars      & \num{0.00034} & \num{1.524}   & \num{0.001}   & \num{0.01} \\
    \midrule
    \multirow{5}{*}{\shortstack[l]{Primordial giants}} 
    	& Jupiter   & \num{1.00000} & \num{5.469}   & \num{0.003} & \num{0.05} \\ 
	    & Saturn    & \num{0.29943} & \num{7.457}   & \num{0.011} & \num{0.02} \\ 
	    & Ice 1     & \num{0.05307} & \num{10.108}  & \num{0.017} & \num{0.11} \\ 
	    & Ice 2     & \num{0.05307} & \num{16.080}  & \num{0.006} & \num{0.07} \\ 
	    & Ice 3     & \num{0.05411} & \num{22.172}  & \num{0.002} & \num{0.05} \\
    \midrule
    \multirow{4}{*}{\shortstack[l]{Present giants}} 
    	& Jupiter   & \num{1.00000} & \num{5.205}   & \num{0.049} & \num{1.30} \\ 
	    & Saturn    & \num{0.29943} & \num{9.581}   & \num{0.056} & \num{2.48} \\ 
	    & Uranus    & \num{0.04573} & \num{19.230}  & \num{0.044} & \num{0.77} \\ 
	    & Neptune   & \num{0.05395} & \num{30.097}  & \num{0.011} & \num{1.77} \\ 
    \midrule
    \multirow{2}{*}{Halfearths}
        & Halfearth 1    & \num{0.00159} or \num{0.00191} & $1.000-\Delta a$   & \num{0.001}   & \num{0.01} \\
    	& Halfearth 2    & \num{0.00159} or \num{0.00127} & $1.000+\Delta a$   & \num{0.001}   & \num{0.01} \\
    \midrule
    \multirow{2}{*}{Marsplus}
        & Earth & \num{0.00314} & \num{1.000}   & \num{0.001}   & \num{0.01} \\
		& Mars+ & \num{0.00034} & \num{1.100} or \num{1.200}   & \num{0.001}   & \num{0.01} \\
	\bottomrule
    \end{tabular}%
\end{table*}%

The paper is organised as follows. In section \ref{methods}, we justify the initial setup and present the methodology applied. The results are analysed and discussed in section \ref{results}. Finally, section \ref{conclusions} is devoted to the conclusions.  

\section{Methods}\label{methods}

In our simulations, we assume a set of models consisting of an initial system of terrestrial planets with five proto-planets. This intends to mimic possible configurations of the Moon-forming impactor that may have existed before the impact. The model configurations include a proto-Mercury, a proto-Venus, and a proto-Mars, with their current masses and at their current heliocentric distances. They also include two proto-planets, that we will refer to as Halfearths (or h$\oplus$), each of them with a fraction of the mass of the current Earth+Moon system. The mass fraction is defined by the parameter $\gamma$ such that:
\begin{equation}
    m_{\,\mathrm{h}\oplus,1}=(1-\gamma)(m_{\oplus}+m_{\small\rightmoon})\,,\qquad
    m_{\,\mathrm{h}\oplus,2}=\gamma(m_{\oplus}+m_{\small\rightmoon})\,.
\end{equation}
The Halfearths are initially located around 1 au, and separated in semi-major axis by a distance $\Delta a$. We consider two values of $\gamma =0.4$ and 0.5, and five values of $\Delta a =0.05$, 0.08, 0.10, 0.15, and 0.20 au. The values of $\gamma$ are based on the values suggested by \citet{Canup2012}. The minimum value of $\Delta a$ is based on the Hill stability criterion, which in this case requires $\Delta a > 0.035$~au \citep[e.g.][]{Giuppone2013}, and the maximum value is based on \citet{Quarles2015}. For all the five terrestrial proto-planets, the eccentricities and inclinations are initially set to give almost circular and almost co-planar orbits. 
In addition, we consider another model configuration in which, instead of two Halfearths, we have a proto-Earth at 1 au and an additional proto-Mars at either 1.1 or 1.2 au. We call this latter the Marsplus configuration. No disk of planetesimals is included in the models.

The system of giant planets is initially constituted by Jupiter, Saturn, and three Neptune-size ice giants. The simulations are carried out in two stages. The first stage, hereafter \textbf{Phase 1}, lasts $\num{10}\, \si{My}$ and involves the jumping Jupiter instability. The second stage, hereafter \textbf{Phase 2}, lasts $\num{90}\, \si{My}$ and involves the evolution of the giant planets after the instability.

\begin{figure*}
\centering
\includegraphics[width=0.3\linewidth]{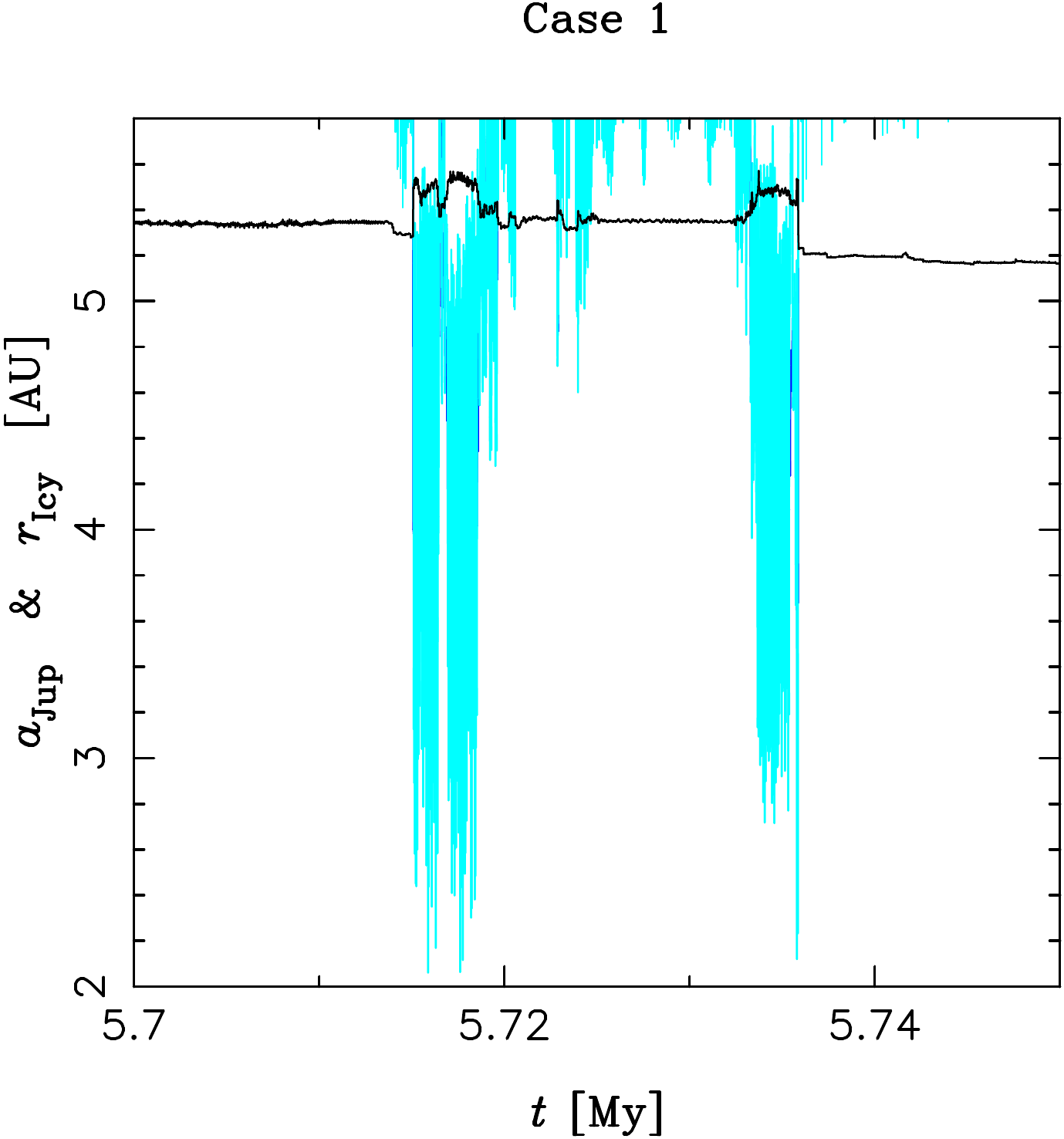}\,\,\,\,
\includegraphics[width=0.3\linewidth]{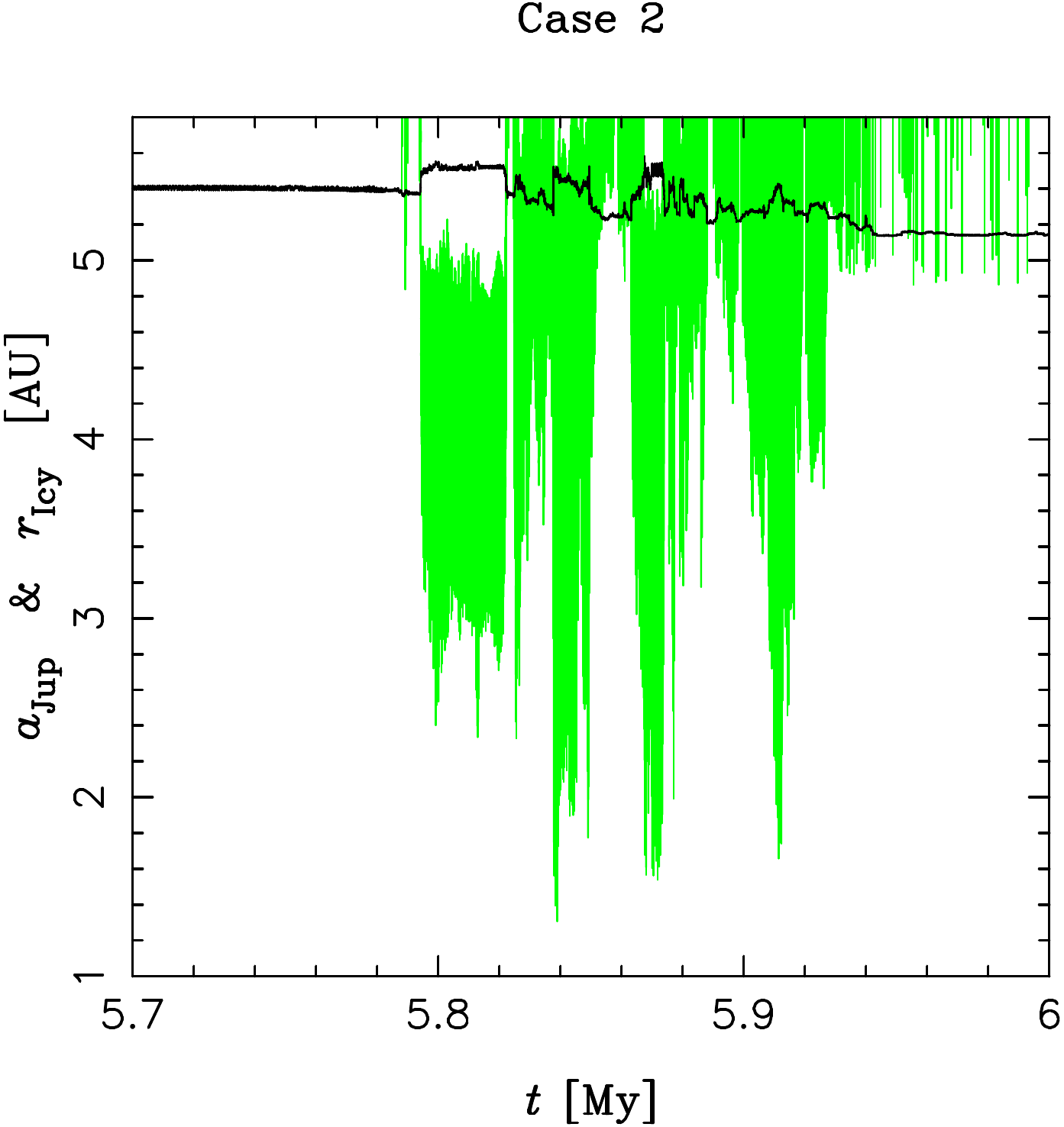}\,\,\,\,
\includegraphics[width=0.3\linewidth]{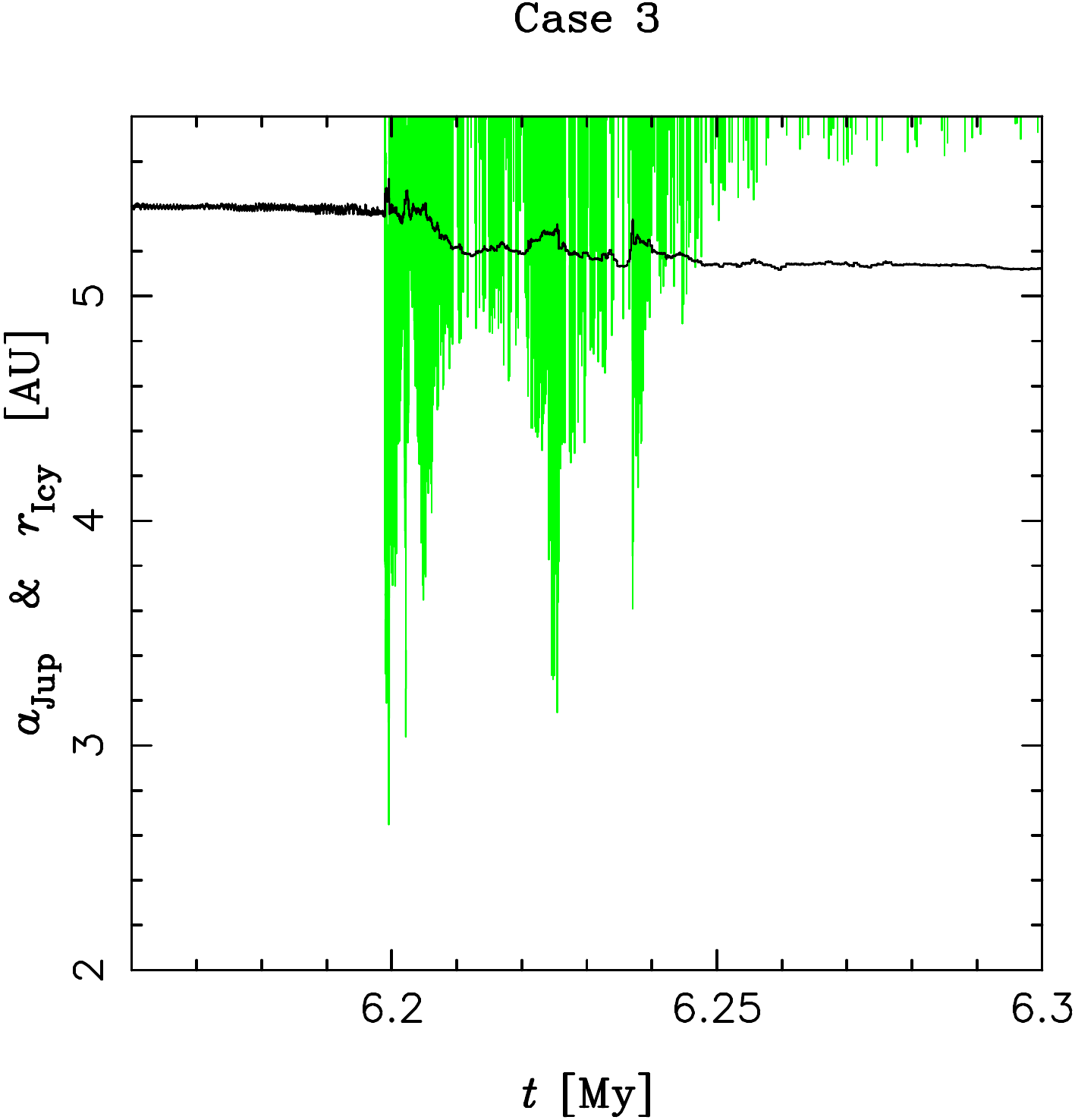}
\caption{Detail of the evolution of Jupiter semi-major axis (black) and the heliocentric distance of the ejected Ice giant (cyan or green, depending on the planet), in the different instability models considered in this study.  We note that the \texttt{CASE1} model involves a fast instability, with few encounters and a ``big'' jump of Jupiter in the end. On the other hand, the \texttt{CASE2/CASE3} models involve much more encounters and a more staggered evolution of Jupiter. Note that the duration of the \texttt{CASE2} instability is $>3$ times longer than in the other cases.}
\label{fig:instcases}
\end{figure*}

During Phase 1, the giant planets evolve following orbits that are already prescribed from previous simulations of the jumping Jupiter instability \citep{Nesvorny2012}. The prescription is generated through the interpolation method described in \citet{Nesvorny2011}. During the instability, Jupiter moves to an inner orbit while one of the ice giants is ejected from the system. We consider three different instability models, labelled \texttt{CASE1}, \texttt{CASE2}, and \texttt{CASE3}. The evolution of the giant planets in each case differs in the number of mutual encounters, the ice giant that is ejected, the precise time when the instability occurs, and the duration of the instability phase. Figure \ref{fig:instcases} shows the evolution of the semi-major axis of Jupiter and the heliocentric distance of the ejected Ice giant during the instability in the three models (see figure caption for details). In particular, the \texttt{CASE1} instability model has provided very successful constraints in previous applications \citep[e.g.][]{Deienno2014,Nesvorny2014,Roig2015,BRASIL2016,Nesvorny2017AJ,Brasil2017,Nesvorny2018AJ}. 

During Phase 2, the giant planets suffer a residual smooth radial migration that leads them to their present heliocentric distances. 
We also perform a few simulations with the giant planets ``fixed'' in the orbits that they ended after Phase 1. These simulations are intended to check if the residual migration may cause an additional effect on the terrestrials, besides that of the jumping Jupiter instability.

In addition to the above models, we also performed simulations with no instability at all, keeping the giant planets in ``fixed'' orbits, i.e. without any migration: 
\begin{enumerate}
    \item one model considers the five giant planets in their primordial orbits (i.e., those at the beginning of Phase 1), in which the giants are placed in a mutual resonant compact configuration, with Jupiter and Saturn in a 3:2 mean motion resonance. We call this the \texttt{JOVINI} model;
    \item another model considers the four giant planets in their present orbits. We call this the \texttt{JOVEND} model.
\end{enumerate}
All the simulations in these models without an instability lasted 100 Myr, covering the same time interval as the Phase 1 + Phase 2 simulations with an instability. These simulations are used as a control group.

Table \ref{tab:initial_conditions} summarises the initial values of the orbital elements considered in the different models and configurations. For each possible configuration of $\Delta a$ and $\gamma$ for the terrestrial planets in the \texttt{CASE1}, \texttt{CASE2}, \texttt{CASE3}, \texttt{JOVINI} and \texttt{JOVEND} models, we perform \num{20} different simulations, that differ in the initial orbital angles (longitude of node, argument of perihelion, and mean anomaly) of the terrestrial planets. These angles are set at random between 0 and 360 deg. In total, we performed more than 1000 simulations.

The simulations are carried out using the N-body symplectic algorithm SyMBA \citep{Duncan1998}. For the simulations considering the giants instability during Phase 1, we used a modified version of this code called iSyMBA (\citealp{Nesvorny2021}; Roig et al. in preparation), that allows to interpolate the positions and velocities of the giant planets from the prefixed orbits that are stored in a file. The five terrestrial proto-planets gravitationally interact with one another and also feel the perturbation from the giant planets, but they do not perturb the giant planets. This approach is similar to the one applied in our previous study of the terrestrial system \citep{Roig2016}, but here the terrestrial proto-planets are allowed to have close encounters and collisions with one another, which are properly manipulated by the symplectic algorithm. 

For the simulations considering the residual migration of the giant planets during Phase 2, we use another version of the SyMBA code, where the smooth radial migration is mimicked by the addition of non conservative forces \citep[see][for details]{Roig2015}. 
This algorithm has been calibrated to also damp the giant planets’ eccentricities to their present values.
For the simulations that consider neither instability nor residual migration, we use the standard version of the SyMBA code.

During the analysis of the results, we focus on the binary collisions among the terrestrial proto-planets. For each collision, we registered the time of the event, the impact velocity, and the impact angle. The impact velocity, $v_{\mathrm{imp}}$ is obtained from the relative velocity of the two bodies, $v_{\mathrm{rel}}$, immediately before the collision:
\begin{equation}
    v_{\mathrm{imp}}=\sqrt{v_{\mathrm{rel}}^2+v_{\mathrm{esc}}^2}\,,
\end{equation}
where
\begin{equation}
    v_{\mathrm{esc}}^2=2G\frac{m_1 +m_2}{R_1 +R_2}\,,
\end{equation}
being $m_i,R_i$ the masses and radii of the two bodies.
The impact angle $\xi$ is related to the impact parameter $b=\sin\xi$, and is computed from the relative position of the bodies, $r_{\mathrm{rel}}$, as:
\begin{equation}
    \sin\xi =\sin\alpha \frac{r_{\mathrm{rel}}}{R_1+R_2}\,,
\end{equation}
with
\begin{equation}
    \cos\alpha =\frac{ \vec{r}_{\mathrm{rel}} \cdot\vec{v}_{\mathrm{rel}} }{ r_{\mathrm{rel}} v_{\mathrm{rel}}}\,.
\end{equation}

In order to compare the final state of the simulations with the present solar system, we use the radial mass concentration (RMC) and the angular momentum deficit (AMD), defined as:
\begin{equation}
    RMC = \max_a\left( \frac{\sum_k m_k}{\sum_k m_k [\log(a/a_k)]^2} \right)\,,\label{RMC}
\end{equation}
\begin{equation}
    AMD = \frac{\sum_k m_k \sqrt{a_k} \left(1-\sqrt{1-e_k^2} \cos I_k \right)}{\sum_k m_k \sqrt{a_k}}\,,
\end{equation}
where $m_k,a_k,e_k,I_k$ are the masses, semi-major axes, eccentricities, and inclinations, the index $k$ runs over the terrestrial planets only, and the maximum in Eq. (\ref{RMC}) is computed over the interval $0.1\leq a\leq 2.1$~au.

\section{Results}\label{results}

\begin{figure}
\centering
\includegraphics[width=\linewidth]{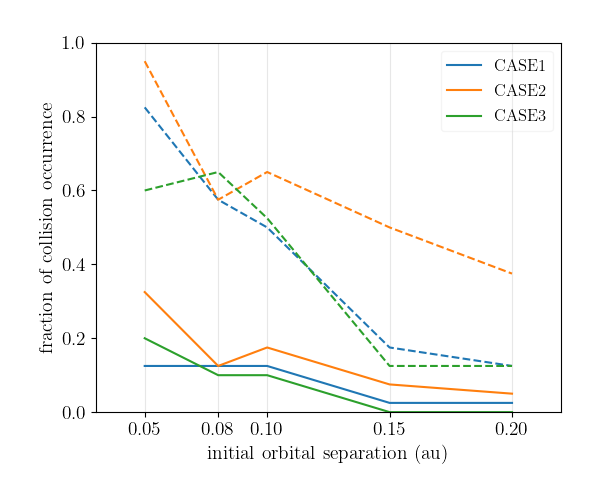}
\caption{The fraction of simulations in the Halfearths models that lead to collisions among the terrestrial planets. The horizontal axis is the initial orbital separation between the two Halfearths. The dashed lines consider all the collisions, and the full lines only the successful cases, i.e. when only one collision between the two Halfearths is recorded and the final system resembles the present inner solar system (see text for details).}
\label{fig:fraccoll}
\end{figure}

\begin{figure}
\centering
\includegraphics[width=\linewidth]{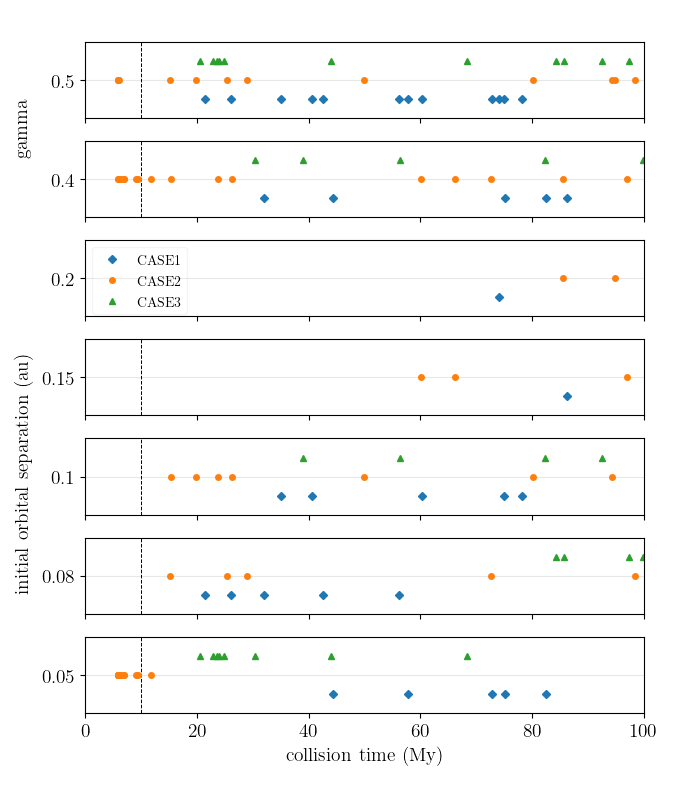}
\caption{The collision times in the successful simulations of the Halfearths models, as a function of the initial separation between the Halfearths (bottom five panels) and of $\gamma$ (top two panels). Different symbols correspond to the different instability models. The vertical dashed line marks the first 10 My of evolution during which the instability occurs (Phase 1).}
\label{fig:timecoll}
\end{figure}

\begin{figure}
    \centering    \includegraphics[width=\linewidth]{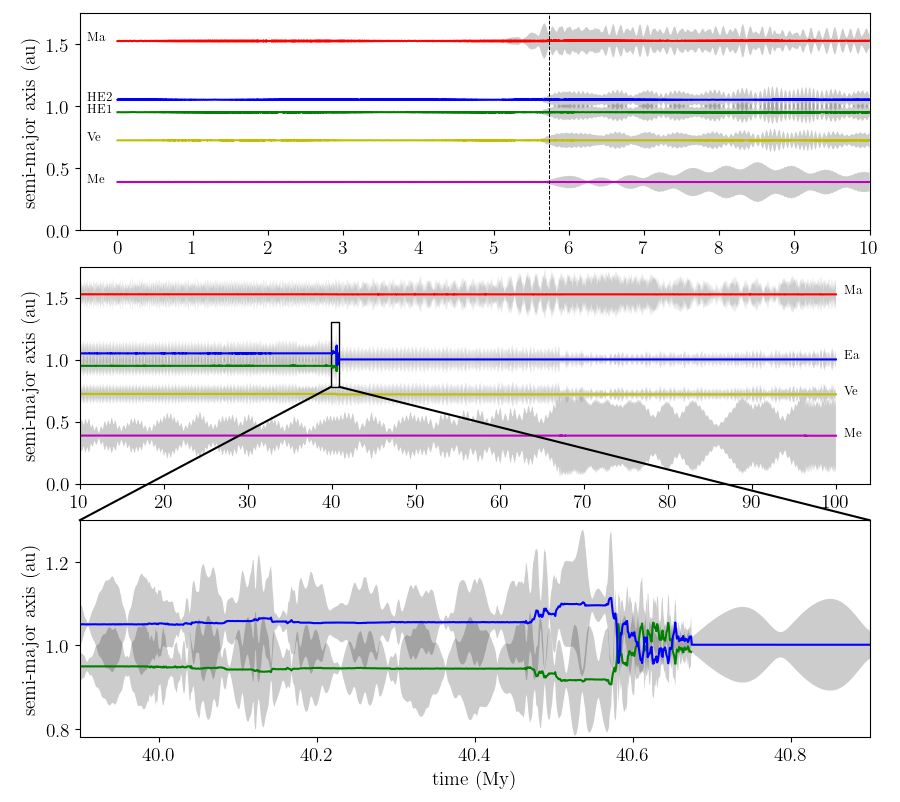}
    \caption[halfearth\_00003 simulation]{An example of a successful Halfearths simulation in the \texttt{CASE1} model, considering $\Delta a = 0.1$ and $\gamma =0.5$. The top panel corresponds to Phase 1, and the middle panel to Phase 2. The bottom panel is a zoom of the collision. The semi-major axes of the five terrestrials are shown (colour lines), together with the range of perihelion and aphelion distances (shaded areas): Mercury (Me - magenta), Venus (Ve - yellow), Halfearth 1 (HE1 - green), Halfearth 2 (HE2 - blue), and Mars (Ma - red). The vertical dashed line in the top panel indicates the time of the instability. The collision happens at about 40.6 My, with a relation $v_\mathrm{imp}/v_\mathrm{esc}=1.414$ and an impact angle $\xi =20.27^\circ$ ($b=0.347$).}
    \label{fig:halfearth-simulation-with-instability-detals}
\end{figure}

\subsection{Halfearths simulations}

The models considering the two Halfearths configuration lead to collisions among the terrestrial planets in 48.5\% of the simulations with an instability of the giant planets. In most cases, only one collision occurs, but 14.5\% of the simulations record up to two collisions, and 0.8\% record up to three collisions. Of particular interest are those simulations recording a collision of the two Halfearths, which represent 20.8\%. Within this latter group, a simulation is considered to be successful if, after the Halfearths collision, the system reaches a final configuration at the end of the simulation that resembles the present inner solar system. This is evaluated in terms of orbital distribution, RMC, and AMD, as discussed below. In particular, we require a final state with 4 terrestrial planets and $RMC \simeq 89$ to classify a simulation as successful. These successful simulations represent 10.5\% of the total simulations. 

The unsuccessful simulations record, in most cases, collisions involving Venus (either Venus with Mercury or Venus with a Halfearth), or collisions involving the two Halfearths that lead to wrong final configurations (usually Mercury is lost by hitting the Sun), or, in very few cases, collisions involving Mars.

On the other hand, in the control simulations without an instability (\texttt{JOVINI} and \texttt{JOVEND}), the giant planets remain all the time in their initial orbits, with normal secular variations. 
The five terrestrial planets maintain their almost circular and co-planar initial configurations (see Table \ref{tab:amdrmc}). No collisions are recorded in these simulations.

Figure \ref{fig:fraccoll} shows the fraction of simulations that record collisions as a function of the initial separation of the Halfearths, for the different models. The dashed lines consider all collisions, and the solid lines correspond to the successful simulations. The instability of the giant planets triggers a lot of collisions, especially in the \texttt{CASE2} model, and seems to be relevant to favour the collisions between the Halfearths. There is also a clear dependence of the number of collisions with the initial separation of the Halfearths. 
A similar analysis in terms of $\gamma$ shows that both the simulations with collisions and the successful ones are $\sim 1.3$ times more frequent for $\gamma=0.5$ than for $\gamma=0.4$. 

In the remaining of this section, we will focus on the analysis of the successful simulations only which, as mentioned before, represent 10.5\% of the total simulations. 

Figure \ref{fig:timecoll} shows the distribution of the collision times between the Halfearths as a function of $\Delta a$ and $\gamma$. The most important result here is that, in spite of the instability occurring early, during Phase 1, the collisions in general occur much later, during Phase 2. 
This happens because the instability excites the eccentricities of the terrestrial planets, allowing them to start to evolve in mutually crossing orbits, but it takes some time until a close encounter effectively results in a physical collision. 
The only exceptions are some simulations of the \texttt{CASE2} model, that occur in Phase 1, immediately after the instability. In general, collisions in the \texttt{CASE2} model tend to happen earlier than in the other models. This would not be surprising, since the \texttt{CASE2} instability lasts longer, and therefore it causes more excitation of the terrestrial orbits. There is also a clear correlation between the average collision time and the initial separation $\Delta a$, but no specific correlation is observed with $\gamma$. 

The above result is of major relevance for the Moon-forming impact.
It implies that the instability may have happened very early after the dissipation of the gas in the proto-planetary disk, and yet have triggered a Moon-forming impact compatible with the time estimated for the Giant Impact from cosmochemical constraints, that is 30 to $60\, \si{My}$ after the dissipation of the gas.

An example of a successful simulation is shown in Figure \ref{fig:halfearth-simulation-with-instability-detals}. This simulation considers the \texttt{CASE1} model, with $\Delta a=0.1$ and $\gamma =0.5$. In the top panel, corresponding to Phase 1, we see the instability starting at $\sim \num{5.7}\, \si{My}$ and causing an excitation of the orbits. In the middle and bottom panels, corresponding to Phase 2, we show a zoom of the moment when the close encounter and the merger between the two Halfearths occur, at $\sim 40.6\, \si{My}$. After the merger, the resulting Earth-mass planet remains at $\num{1}\,\si{au}$. 
\begin{figure*}
\centering
\includegraphics[width=0.49\linewidth]{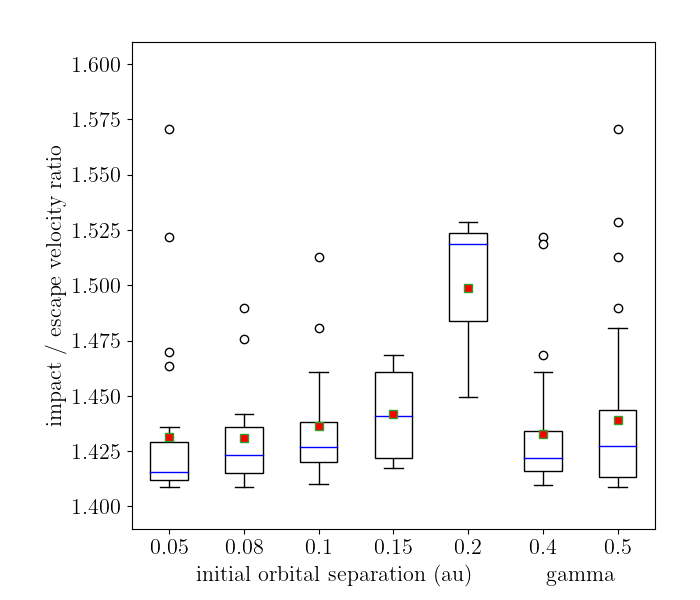}
\includegraphics[width=0.49\linewidth]{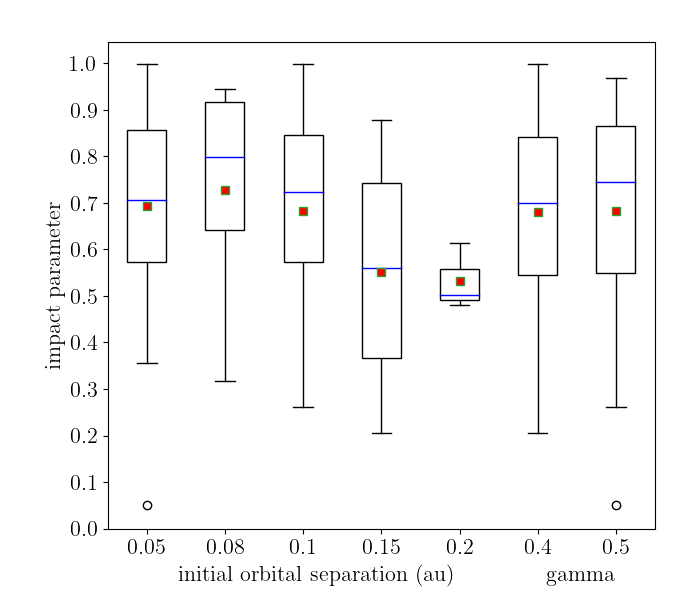}
\caption{Box plot of the distribution of $v_\mathrm{imp}/v_\mathrm{esc}$ ratios (left) and impact parameters (right), as a function of $\Delta a$ and $\gamma$, for the successful simulations of the Halfearths models. The three instability models are considered together. The blue lines indicate the median and the red squares indicate the mean. Open circles represent outliers.}
\label{fig:boxplot}
\end{figure*}

\begin{figure}
    \centering
    \includegraphics[width=\linewidth]{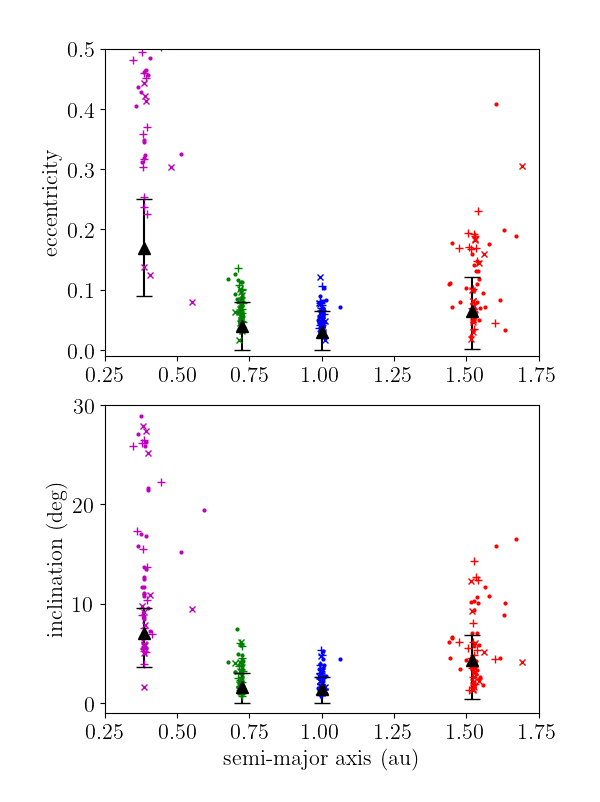}
    \caption[Final eccentricities of the terrestrial planets in the Halfearth group]{Final state of the successful simulations in the Halfearth configuration. Each instability model is identified by crosses (\texttt{CASE1}), dots (\texttt{CASE2}), and plus (\texttt{CASE3}). The colour identify the different planets. The values of the semi-major axis, eccentricity and inclination correspond to the averages over the last 5 My of evolution (during Phase 2). Black triangles indicate the present values, and the error bars correspond to the present minimums and maximums of the secular variations over 5 My. The vertical axes has been truncated for clarity, but Mercury may reach values of $e$ up to 0.75 and $I$ up to $40^\circ$.}
    \label{fig:halfearth-eccentricity}
\end{figure}

The analysis of the impact velocities and angles of the collisions in the successful simulations reveals that the typical ratio $v_\mathrm{imp}/v_\mathrm{esc}$ is around 1.45, and the typical angle $\xi$ is around $50^\circ$. We do not observe any correlation of the distribution of impact velocities and angles with the instability models, except for the fact that in the \texttt{CASE3} model, the velocity distribution has a tail at slightly larger values. Thus, in the following analysis, the three instability models are considered together. 

Figure \ref{fig:boxplot} shows the distributions of the velocity ratio and impact parameter $b$ as a function of $\Delta a$ and $\gamma$. The median values lie in the range 1.42-1.52. For the impact parameter, the median values lie between 0.5-0.8. These values shall be compared to those obtained by \cite{Canup2012}, who determined that the most likely values to produce a ``good'' Moon-forming impact are between 1.0-1.6 for $v_\mathrm{imp}/v_\mathrm{esc}$, with a median of 1.1, and between 0.35-0.70 for $b$, with a median of 0.55. All our values of $v_\mathrm{imp}/v_\mathrm{esc}$ are above 1.4, implying that our collisions are, in general, more energetic that those of Canup. On the other hand, the impact parameter shows a wider distribution, that overlaps with Canup's interval. In particular, 1.7\% of all the simulations (15.9\% of the successful simulations) led to collisions with $v_\mathrm{imp}/v_\mathrm{esc}\simeq 1.4$ and $b\simeq 0.4$, that are compatible with one of Canup's simulations. The simulation shown in Figure \ref{fig:halfearth-simulation-with-instability-detals} is precisely one of these cases. It is worth noting the correlation between the impact velocities and the initial separation, and the anti correlation between the impact parameter and the initial separation. This may imply that there is some subtle anti correlation between the impact velocity and parameter, that would be compatible with Canup's results.

A comparison of the impact velocities and parameters with the results of \citet{Leinhardt2012} indicate that most of the successful collisions (79\%) fall in the hit-and-run regime ($b\gtrsim 0.5$), with partial erosion of the projectile. This regime is not compatible with the Moon's formation impact from two half-Earths. However, 21\% of the successful collisions fall in the partial accretion regime, that may lead to the desired result (as mentioned above, 15.9\% are indeed compatible with Canup's configurations). Notwithstanding, all the collisions are above the merging threshold ($v_\mathrm{imp}/v_\mathrm{esc} \simeq 1$), implying that the inelastic accretion forced by the SyMBA algorithm may be unrealistic, and the final results must be taken only as a first order approximation to the real problem.

A summary of the final states of all the successful simulations of the Halfearth configuration with instability is presented in Fig. \ref{fig:halfearth-eccentricity}, and Table \ref{tab:amdrmc}. We can see that, in general, the eccentricity of Mercury is much more excited than the current value, a fact that has been already addressed in \citet{Roig2016}. In that paper, we showed that the excitation could be amended by including relativistic corrections to the perihelion precession frequency, $g_1$. In this work, reproducing Mercury’s eccentricities and inclinations was not the focus, and since relativistic effects do not significantly affect the other terrestrial planets, they were not included in the simulations. 
Nevertheless, based on our previous result, we may expect that relativistic corrections will decrease the final eccentricities of Mercury by a factor of $\sim 2$.

In Figure \ref{fig:halfearth-eccentricity}, we also note that, in general, Mars ends with the right values of orbital eccentricity and inclination. This differs from the results of \citet{Roig2016}, where Mars' orbit was under-excited, particularly in terms of inclination, which was actually a problem. We must bear in mind, however, that \citet{Roig2016} considered a system with only four terrestrial planets and did not account for the evolution of the system during Phase 2.

Table \ref{tab:amdrmc} reports the values of RMC and AMD for the successful simulations. As expected, the RMC is well constrained and the mean is close to the current inner solar system value of 89.9. On the other hand, the AMD is much less constrained, and its value is quite large compared to the current value of 0.0018. This discrepancy is expected in view of the significant excitation observed in Fig. \ref{fig:halfearth-eccentricity}, especially in the case of Mercury, that may reach values of $e$ up to $\sim 0.75$ and $I$ up to $40^\circ$. Only 6.3\% of the successful simulations (0.7\% of all the simulations) show values of AMD similar to the present one. We do not find any correlation of AMD, nor RMC, with either $\Delta a$ or $\gamma$, but the largest values of AMD are obtained for $\Delta a=0.2$~au.

The excess of AMD in our simulations might be a consequence of the simplifications of the model. Including relativistic corrections, as mentioned above, or the interaction of the proto-planets with a swarm of planetesimals, may lead to different results.

Table \ref{tab:amdrmc} also reports the final values of RMC and AMD for the control simulations. We verify that the final AMD values ($\sim 10^{-4}$) are typically two orders of magnitude larger than the initial ones ($5\times 10^{-7}$). We do not find any systematic dependence of the final values on the initial configuration. 

\begin{table}
  \centering
  \caption{The mean and standard deviation of the final AMD and RMC distributions of the inner planets, for the 10.5\% successful simulations of the Halfearths configuration, in the different instability models. We also give the values corresponding to the control simulations without an instability; in this case the reported AMD is an upper limit.}%
  \label{tab:amdrmc}%
    \begin{tabular}{lcccc}%
    \toprule
      & $\left< RMC\right>$ & $\sigma_{RMC}$ & $\left< AMD\right>$ & $\sigma_{AMD}$ \\
     \texttt{CASE1} & 89.24 & 3.51 & 0.00702 & 0.00302 \\
     \texttt{CASE2} & 88.19 & 7.24 & 0.00895 & 0.00404 \\
     \texttt{CASE3} & 89.61 & 4.48 & 0.00682 & 0.00306 \\
     \texttt{JOVINI} & 88.11 & 0.20 & $<0.00015$ & -- \\
     \texttt{JOVEND} & 88.20 & 0.12 & $<0.00009$ & -- \\
	\bottomrule
    \end{tabular}%
\end{table}%

\subsection{Marsplus simulations}

\begin{figure}
\centering
\includegraphics[width=\linewidth]{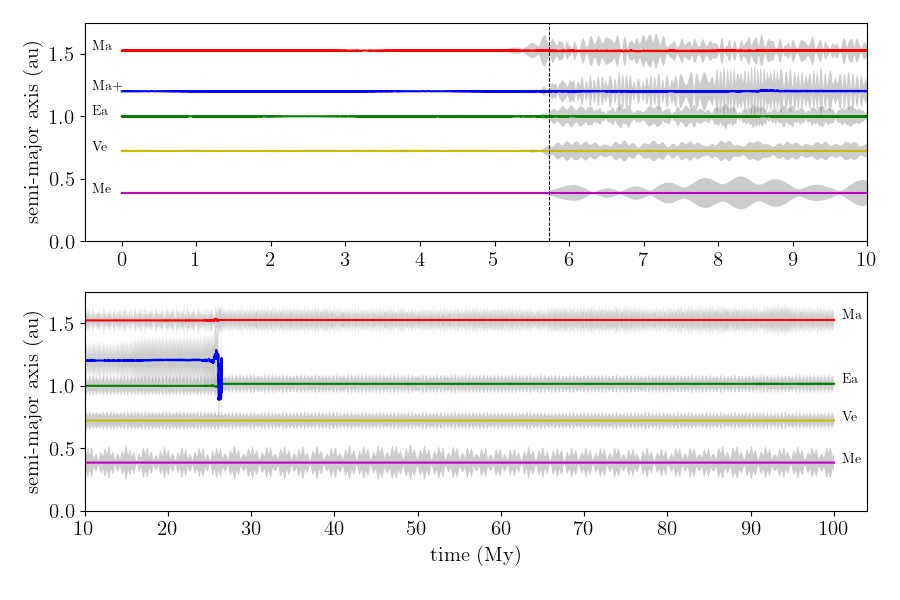}
\caption[mars\_00007 simulation]{An example of a successful evolution in the Marsplus configuration, considering the \texttt{CASE1} instability model and an initial separation $\Delta a =0.2$ between the Earth and the Mars+ planets. The top panel corresponds to Phase 1, and the bottom panel to Phase 2. Colour codes are the same as in Fig. \ref{fig:halfearth-simulation-with-instability-detals}.}
\label{fig:mars1.2_00007_unstable}
\end{figure}

The Marsplus configuration has been tested only in the \texttt{CASE1} instability model. This configuration leads to collisions among the terrestrial planets in 50\% of the simulations, with 20\% of simulations recording two or more collisions. Only 7.5\% of the simulations result in successful cases, i.e. only one collision between the Earth and the additional Mars-like planet, and a final state of the system compatible with the current inner solar system. One of these successful examples is shown in Figure \ref{fig:mars1.2_00007_unstable}. 
Comparing these percents with those of the Halfearths configuration, for the \texttt{CASE1} instability model, we may conclude that the Marsplus configuration performs worse than the Halfearths configuration, because it generates about 25\% fewer successful collisions. However, the statistics in the Marsplus configuration are biased by the smaller number of simulations performed, and would deserve a more exhaustive analysis in a future work. 

It is worth noting that the collisions in the Marsplus configuration also happen exclusively during Phase 2, and they are 4 times more frequent when the initial separation between the Earth and the Mars+ is 0.2 au. We also observe that, in the final state of the system, Mercury ends less excited and the current Mars ends more excited than in the Halfearths simulations.
In the Marsplus configuration, the \texttt{JOVINI} and \texttt{JOVEND} control simulations do not record any collision either. 

\section{Conclusions}\label{conclusions}

Our results indicate that the instability of the giant planets in the jumping-Jupiter model could help to trigger collisions between proto-terrestrial planets, and this may eventually lead to a Moon-forming impact. Collisions occur in about 50\% of the simulations where the instability is considered, compared to no collisions when the instability is not included. 

However, collisions involving a proto-Earth that lead to a successful evolution, in that the final structure of the terrestrial system resembles the present architecture, are not too frequent and occur in only 10\% of the simulations. 

The initial configuration with two proto-Earths, each with approximately half mass of the current Earth, produces more such successful cases than the configuration considering a proto-Earth and an additional proto-Mars.

We find that there is a significant delay of more than $\sim 20\,\si{My}$ between the time of the instability and the occurrence of the collisions. This result is of most relevance for the Moon-forming impact. It implies that the instability may have happened very early after the dissipation of the gas in the proto-planetary disk, and yet have triggered a Moon-forming impact at a time compatible with those estimated from cosmochemical constraints (30 to $60\, \si{My}$ after the dissipation of the gas). 

While the timing of the Moon-forming impact is still very much debated, for the purposes of what has been investigated here, changing the timing by 10s of My should not matter, provided that the Earth-Impactor configuration is somewhat stable, as demonstrated in our control runs.

We also find that the jumping Jupiter instability is the main trigger for the Giant Impact, and our results are independent of the modelled post-instability migration scheme. In other words, during the Phase 2 of the simulations, the results considering the giant planets in fixed orbits after the instability are indistinguishable from those considering a smooth radial migration.

An analysis of the collision geometry in the successful simulations shows that, in general, the recorded impacts lie in the hit-and-run regime. However, about 20\% of the successful simulations occur in the regime of partial accretion, and about 15\% would be compatible with the geometries predicted by \citet{Canup2012} for a Moon-forming impact. 

The final values of the RMC in the successful simulations is consistent with the current value for the inner solar system, meaning that the planets ended at the correct heliocentric distances. However, the values of the AMD are about 4 times larger than the present value,on average. This is mostly due to the fact that Mercury's orbit is significantly excited, especially in terms of eccentricity. Only 6\% of the successful simulations produce values of RMC and AMD similar to the present ones.

Analysing the dependence of the successful results on the initial configurations of the model with two half-Earths, we verify that the best results occur for an initial radial separation of 0.1 au. Smaller and larger separations produce results that are not good either in terms of collision geometry, or final AMD, or both. 

The results also show a dependence on the instability model. If the instability lasts some 100~ky, the terrestrial planets become more excited and collisions are triggered earlier. Also, the final AMD is slightly larger in this case. More delayed collisions and smaller values of AMD are obtained with instabilities that last $\lesssim 50$~ky. There is no dependence of the impact geometries with the instability model.

In the end, we must admit that the fraction of very successful cases in our simulations is quite small. This leads us to conclude that the jumping Jupiter instability is a plausible mechanism to trigger the Moon-forming impact, but maybe not an efficient one. 

While giant impacts are ubiquitous in classic simulations of terrestrial planet formation, we must recall that our models assume that the terrestrial proto-planets are already formed, or almost formed, when the instability happens. This can be justified in view of recent results supporting the fast formation of the terrestrial planets, while gas was probably still present in the proto-planetary disk \citep{Ogihara2018}. High-resolution simulations \citep{Walsh2019,Clement2020} find that quasi-stable systems of more massive proto-planets may emerge from the gas disk and struggle to accrete. Thus, conditions akin to the ones tested in this study might be somewhat realistic.

On the other hand, a critical point of our simulations is that the initial conditions of the terrestrial planets in quasi circular and quasi co-planar orbits are likely unrealistic. Mechanisms like dynamical friction \citep[][]{OBrien2006,Raymond2006} and collisional fragmentation \citep[][]{Chambers2013,Clement2019} may help to damp the eccentricities and inclinations, but the initial AMD of our simulations is extremely low and inconsistent with what comes out of terrestrial planet formation models. Nevertheless, our aim was to address the effects of the giant planets instability using a sort of toy model, with no pretensions of being realistic. Besides, working with a more realistic scenario would face the problem that initial conditions are uncertain, and in the end we do not know what was the actual primordial AMD of the inner solar system.

Finally, simulations in the framework of the standard model of terrestrial planets formation, including, for example, the interaction of the proto-planets with a disk of remnants planetesimals, may lead to different results than the ones presented here, and it is worth of analysis in a future study.

\section*{Acknowledgements}
We wish to thank the referee Matt Clement for his very helpful and constructive review.
The simulations included in this work have been performed at  the  SDumont cluster  of  the  Brazilian  System  of High-Performance Computing (SINAPAD). The authors acknowledge support form the Brazilian National Council of Research (CNPq).

\section*{Data availability}
The data underlying this article will be shared on reasonable request to the corresponding author.



\input{moonformation-rev3.bbl}

\bsp	
\label{lastpage}
\end{document}